\documentclass[12pt]{iopart}
\usepackage{amssymb}
\usepackage[amssymb]{SIunits}
\usepackage{graphicx}
\usepackage{xspace}
\usepackage{bm}
\usepackage[colorlinks,urlcolor=blue]{hyperref}
\usepackage[all]{hypcap}
\newcommand*{\doi}[1]{\href{http://dx.doi.org/\detokenize{#1}}{\detokenize{#1}}}

\begin{document}

\title[Magnetic mesh generation and field line reconstruction for SOL and divertor modeling in stellarators \,]{Magnetic mesh generation and field line reconstruction for scrape-off layer and divertor modeling in stellarators}

\author{H. Frerichs${}^1$, D. Boeyaert${}^1$, Y. Feng${}^2$, K.A. Garcia${}^1$}

\address{${}^1$ Department of Nuclear Engineering \& Engineering Physics, University of Wisconsin - Madison, WI, USA}
\address{${}^2$ Max-Planck-Institut f\"ur Plasmaphysik, Greifswald, Germany}

\ead{hfrerichs@wisc.edu}

\begin{abstract}
The design of divertor targets and baffles for optimal heat and particle exhaust from
magnetically confined fusion plasmas requires a combination of fast, low-fidelity models
(such as EMC3-Lite \cite{Feng2022}) for scoping studies and high-fidelity ones (such as EMC3-EIRENE
\cite{Feng2014}) for verification. Both of those approaches benefit from a magnetic flux tube mesh
for fast interpolation and mapping of field line segments \cite{Feng2005}. A new automated mesh
generator for unstructured quadrilateral flux tubes with adaptive refinement is presented and integrated into FLARE \cite{Frerichs2024a}.
For HSX with an extended first wall, it is found that several layers of flux tubes can
span the entire half field period before splitting is required. This is an advantage over the
traditional setup of the EMC3-EIRENE mesh where careful construction of several sub-
domains is required already for the much tighter present first wall.
In particular, there is no longer the need to manually construct a suitable outer boundary for the mesh.
The divide and conquer paradigm with unstructured mesh layout offers a powerful alternative
for fast head load approximation that is suitable for integration into optimization workflows.
Further examples for W7-X and CTH demonstrate the versatile application range.
\end{abstract}

\vspace{2pc}

\def\vec#1{\ensuremath{{\bf #1}}\xspace}
\def\Poincare{Poincar\'e\xspace}
\def\phimap{\ensuremath{\varphi_{\textnormal{map}}}\xspace}
\def\Dmax{\ensuremath{\Delta_{\Phi \, \textnormal{max}}}\xspace}
\def\Davg{\ensuremath{\Delta_{\Phi \, \textnormal{avg}}}\xspace}
\def\mpadapt{\ensuremath{m_p^{(j)}}\xspace}
\def\mpnecessary{\ensuremath{m_p^{\circ}}\xspace}
\def\qproxy{\ensuremath{q_{t \, (\chi_\parallel)}}\xspace}
\def\PSOL{\ensuremath{P_{\mathrm{SOL}}}\xspace}

\section{Introduction}

The stellarator path towards magnetic confinement fusion power offers great design flexibility (e.g. number of field periods, modular vs. helical coils, quasi-symmetric vs. quasi-isodynamic configurations ...) for leveraging its advantage of improved stability and intrinsic steady-state operation compared to tokamaks \cite{Boozer2015a, Boozer2024}.
A common challenge for both concepts is the control of particle and heat exhaust from the scrape-off layer (SOL) at acceptable levels for plasma material interactions and impurity accumulation \cite{Koenig2002}.
However, the increased complexity of stellarators introduces further difficulties for divertor design such as toroidal localization of heat load peaks and inefficient pumping of neutral particles due to poor closure \cite{Feng2021, Boeyaert2024}.

The optimization of divertor targets and baffles requires a combination of fast, low-fidelity models for scoping studies and high-fidelity ones for verification.
The EMC3-EIRENE code \cite{Feng2004, Reiter2005, Feng2014} is currently the workhorse for divertor and scrape-off layer analysis in W7-X \cite{Feng2016, Effenberg2019, Feng2021, Schmitz2021, Boeyaert2024}.
It is based on a Monte Carlo fluid model for the steady state plasma (EMC3) which is coupled to a kinetic model for neutrals (EIRENE), and it includes cooling from radiation of intrinsic or seeded impurities.
Self-consistent solution of the non-linear equations requires an iterative procedure, and convergence can take a significant number of iterations under detached conditions.
Therefore, it is currently not suitable to integrate EMC3-EIRENE into an optimization workflow.
Nevertheless, a linearized reduced model (EMC3-Lite \cite{Feng2022}) has been developed recently for fast approximation of heat loads in support of the design of a new divertor at W7-X \cite{Gao2023}.
Even though EMC3-Lite does not include detachment physics, it can be used to minimize the heat loads that a divertor needs to handle in the first place.

Both of these approaches benefit from a magnetic flux tube mesh for fast interpolation and mapping of field line segments \cite{Feng2005}.
However, construction of such a mesh often needs manual adjustments in order to avoid gaps in the simulation domain and to ensure that the cross-sections of all flux tubes remain convex in order to support a unique inverse transformation from cylindrical to local coordinates (see cartoon of good vs. bad flux tube in figure \ref{fig:finite_flux_tube}).
The 3D mesh is typically constructed by field line tracing from a structured 2D base mesh (several toroidal blocks may be required in some situations, and a block-structured mesh has been implemented for tokamak configurations \cite{Frerichs2010}).
However, if the base mesh does not extend sufficiently beyond plasma exposed surfaces, gaps can occur where field lines travel too far radially inwards.
On the other hand, proximity to coils and strong magnetic shear can result in flux conservation errors and invalid flux tube cross-sections.
Thus, construction of the base mesh often requires fine tuning.
A quasi-orthogonal mesh may be favorable for good flux tube shape, but can result in extremely packed or inadequate poloidal spacing.
Instead, an unstructured mesh with adaptive refinement may offer a path to avoid these issues.

In a different approach, an adaptive mesh refinement method has been added to SOLEDGE3X-HDG \cite{Piraccini2022} which is based on a high-order hybrid discontinuous Galerkin (HDG) method.
Another alternative to reversible field line mapping is the flux-coordinate independent (FCI) approach, which is e.g. applied by GRILLIX \cite{Stegmeir2017} and Hermes-3 \cite{Dudson2024}.
Recent advances in that approach allow for a finite difference based discretization of parallel operators with high accuracy, which is required in order to avoid that numerical perpendicular diffusion dominates the slow perpendicular transport in a mesh that is not aligned with flux tubes.
Nevertheless, this approach is still based on a field line map.
The difference, however, is that field lines are only traced up to the neighboring poloidal planes.
In particle based (Monte Carlo) methods such as EMC3, on the other hand, a large number of trajectories are typically followed, an those have to be mapped from one flux tube to the next.
Therefore, it is more efficient to maximize the local flux tube length here.

A new mesh generator for unstructured quadrilateral flux tubes with adaptive refinement is presented in section \ref{sec:adaptive_flux_tubes} and integrated into the FLARE code \cite{Frerichs2024a}.
First, starting from a pair of good flux surfaces, an initial layer of flux tubes is constructed across the specified toroidal domain (such as the half field period in stellarators).
Then, layers of flux tubes are added incrementally while the poloidal resolution can be increased locally as needed.
Whenever the new flux tubes fail quality checks, they are discarded and the current toroidal domain is split in half from there on radially outward.
The entire volume up to the first wall (or any poloidally and toroidally closed magnetic field line casing) is eventually filled with flux tubes.
Divertor targets and baffles can be added and modified later for heat and particle exhaust optimization studies within the given field line casing.
Examples for CTH, HSX and W7-X are presented in section \ref{sec:examples}.


\section{Adaptive flux tube mesh construction} \label{sec:adaptive_flux_tubes}

The common basis for the new mesh generator and the EMC3 standard is a pair of good flux surfaces at or just inside the last closed flux surface.
These flux surfaces can be selected from a \Poincare plot from which a smooth representation can be generated (see e.g. section 3.5.2 in \cite{Frerichs2024a}).
Typically, the toroidal domain of the simulation is taken to be one half field period in stellarators.
For W7-X (with 5 field periods), a single block of 36 deg is usually sufficient.
However, many other applications require that the toroidal domain is split into toroidal blocks, and a base mesh is constructed for each block.
In particular, the simulation domain for HSX (with 4 field periods) needs to be split into 5 blocks of 9 deg \cite{Bader2013} due to deformation of flux tubes from large radial excursions of field lines.
Furthermore, the outer boundary for the base mesh in each block needs to be chosen carefully such that no gaps develop in the 3D grid.
This can take several manual iterations, and extending the first wall with space for divertor targets and baffles only further complicates mesh construction \cite{Boeyaert2024a, Schmitt2025}.
In the new approach outlined here, we start from the inner boundary with a single base mesh at the center of the domain and work outwards from there with adjustments as needed.
We will give a brief introduction of field line reconstruction in section \ref{sec:interpolation} before presenting the new algorithm for adaptive flux tube mesh construction in section \ref{sec:algorithm}.

\subsection{Field line reconstruction} \label{sec:interpolation}

\begin{figure}
\begin{center}
\includegraphics[width=80mm]{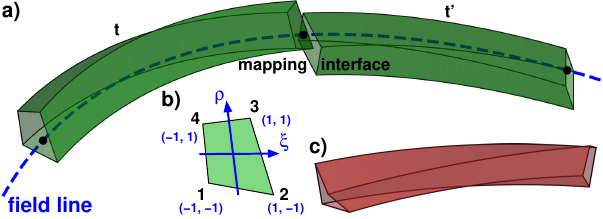}
\caption{(a) Finite flux tube ($\vec{t}$) and an adjacent one ($\vec{t}'$) with a common field line. Drawing is based on figure 2 in \cite{Feng2005}. (b) Local coordinate system $(\xi, \rho)$ within a flux tube cross-section and orientation of guiding field lines. (c) Finite flux tube with invalid (non-convex) cross-section on the left side.}
\label{fig:finite_flux_tube}
\end{center}
\end{figure}

In toroidal configurations, a magnetic field line is determined by integrating the following system of ordinary differential equations

\begin{equation}
\frac{dr}{d\varphi} \, = \, r \, \frac{B_r}{B_\varphi}, \qquad
\frac{dz}{d\varphi} \, = \, r \, \frac{B_z}{B_\varphi} \label{eq:ODE}
\end{equation}

\noindent from an initial point $\vec{p}_i \, = \, \left(r_i, z_i, \varphi_i\right)$ given in cylindrical coordinates.
The right hand side of (\ref{eq:ODE}) is given by the local magnetic field vector $\vec{B}(\vec{p}) \, = \, \left(B_r, B_z, B_\varphi\right)$ at $\vec{p} \, = \, \left(r, z, \varphi\right)$.
In the following, $\vec{F}(\varphi) \, = \, (r(\varphi), z(\varphi))$ refers to a field line segment over some toroidal domain $T$ obtained by numerical integration (see section 3.1 in \cite{Frerichs2024a} for a discussion of numerical methods implemented in FLARE).

Field line reconstruction is based on a mesh of magnetic flux tubes \cite{Feng2005}.
A finite quadrilateral flux tube is defined by four field line segments (see figure \ref{fig:finite_flux_tube}).
Within each flux tube, bilinear interpolation

\begin{equation}
\vec{F}_{\xi \rho}^{(\vec{t})}(\varphi) \, = \, \sum_{i \, = \, 1}^{4} \, \vec{F}_{i \vec{t}}(\varphi) \, \frac{1}{4} \, \left(1 \, + \, \xi_i \, \xi\right) \, \left(1 \, + \, \rho_i \, \rho\right), \quad \varphi \in T \label{eq:interpolation}
\end{equation}

\noindent is applied for given local coordinates $(\xi, \rho) \in [-1,1]^2$.
The coefficients $(\xi_i, \rho_i)$ in (\ref{eq:interpolation}) are the local coordinates associated with the 4 guiding field lines $\vec{F}_{i \vec{t}}$ for flux tube $\vec{t}$ as shown in figure \ref{fig:finite_flux_tube} (b).
Reconstruction of a field line beyond the domain $T$ is based on a reversible mapping

\begin{equation}
\vec{F}_{\xi \rho}^{(\vec{t})}(\phimap) \, = \, \vec{F}_{\xi' \rho'}^{(\vec{t}')}(\phimap)
\label{eq:mapping}
\end{equation}

\noindent at the interface \phimap between flux tube $(\vec{t})$ and an appropriate neighbor $(\vec{t}')$.
More details about the reversible field line mapping can be found in \cite{Feng2005}.
The cross-section of the flux tube must remain convex throughout $T$ such that the inverse transformation from cylindrical to local coordinates is unique.
A counter example is shown in figure \ref{fig:finite_flux_tube} (c) which must be avoided during construction of a magnetic flux tube mesh.


\subsection{New algorithm for mesh construction} \label{sec:algorithm}

\begin{figure}
\begin{center}
\includegraphics[width=80mm]{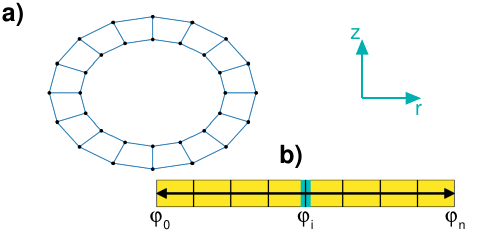}
\caption{(a) Initial layer of cells constructed from pair of good flux surface contours in the r-z plane at the center of the toroidal domain at $\varphi_i$, (b) flux tube segments are constructed by tracing field lines from the base mesh (cyan) towards either end of the toroidal domain.
Yellow color refers to a full length flux tube here and in figures \ref{fig:uqmesh} and \ref{fig:examples}.}
\label{fig:uqmesh_base}
\end{center}
\end{figure}

First, a single layer of cells is constructed from a pair of good flux surface contours at the center of the simulation domain at $\varphi_i$ as indicated in figure \ref{fig:uqmesh_base} (a).
The poloidal resolution $m_p$ is an input parameter and determines the poloidal reference length $\Delta_p$.
Another input is the toroidal discretization $(\varphi_k), k = 0,\ldots,n$.
An initial ring of flux tubes is then generated from field line tracing across the domain $[\varphi_0, \, \varphi_{n}]$.
Because it is generated from good flux surfaces, it can be assumed that this initial ring of flux tubes passes the quality check for flux conservation and shape.
This is indeed the case for the examples presented later.
From here, layers of flux tubes are added incrementally until the entire specified volume is filled with flux tubes.
Divertor targets and baffles are not taken into account at this point.
Rather, a maximal volume is specified in which divertor target and baffle geometry can be optimized later.
This magnetic field line casing should exclude any coils and is required as input.

\begin{figure}
\begin{center}
\includegraphics[width=80mm]{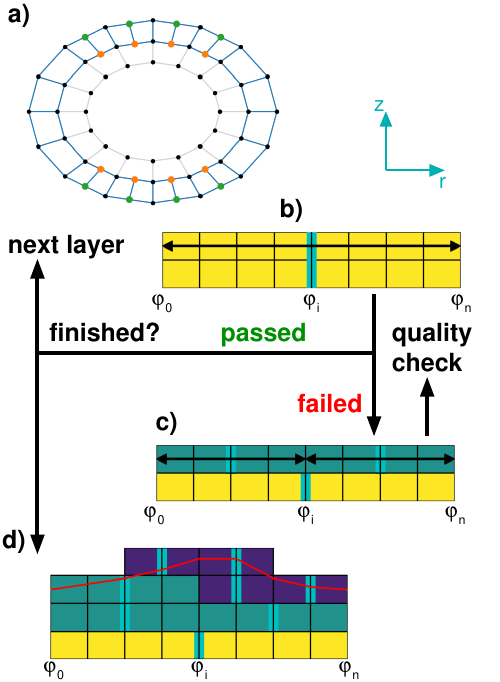}
\caption{(a) New layer of cells (blue) in the r-z plane added to the base mesh with adaptive poloidal refinement, (b) flux tube segments are constructed by tracing field lines from the new layer of the base mesh (cyan) across the toroidal domain, (c) if quality check fails, the domain is divided and a new base mesh is initialized for each sub-domain for which flux tubes are generated (teal color indicates half-length flux tubes here and in figure \ref{fig:examples}), (d) layers are added incrementally until the entire volume inside a magnetic field line casing (red) is filled. Dark purple color indicates minimal-length flux tubes here and in figure \ref{fig:examples}.}
\label{fig:uqmesh}
\end{center}
\end{figure}

A new layer of cells is added to the base mesh as indicated in figure \ref{fig:uqmesh} (a).
For each mesh node $\vec{x}_{ij}$, a new mesh node is placed at

\begin{equation}
\vec{x}_{i \, j+1} \, = \, \vec{x}_{ij} \, + \, \Delta_r^{(j)} \, \vec{r}_{ij},
\qquad
\Delta_r^{(j)} \, = \, (1 + \delta_r) \, \Delta_r^{(j-1)}
\end{equation}

\noindent where $\vec{r}_{ij}$ is the unit normal vector associated with the secant $\vec{x}_{i+1 \, j} \, - \, \vec{x}_{i-1 \, j}$.
The initial radial width $\Delta_r^{(0)} = \Delta_r$ and the relative increment $\delta_r$ are input parameters.
The poloidal resolution \mpadapt of the new ring is determined by $\Delta_p$: new nodes (highlighted green) are inserted in cells where the new (upper) poloidal edge exceeds $2 \, \Delta_p$, and virtual nodes (highlighted orange) are placed along the corresponding old (lower) poloidal edge.
Then, as indicated in figure \ref{fig:uqmesh} (b), flux tubes are generated by tracing field lines from the new layer of nodes in the base mesh across the toroidal domain.
No field lines are traced from virtual nodes.
Rather, a virtual field line is defined from the midpoints of two field lines $i_1$ and $i_2$:

\begin{equation}
\vec{F}_{(i_1, i_2)}(\varphi) \, = \, \frac{1}{2} \, \left(\vec{F}_{i_1}(\varphi) \, + \, \vec{F}_{i_2}(\varphi)\right). \label{eq:bisect}
\end{equation}

\noindent This way, the virtual field line is always located along the straight edge between the two guiding field lines such that no gap is introduced along the way.
The bilinear interpolation (\ref{eq:interpolation}) reduces to a linear interpolation on any of the four sides of a flux tube, and this in turn reduces to (\ref{eq:bisect}) for the midpoint between each pair of guiding field lines.
This also implies a linear equation for the transformation of the local coordinate from the coarser tube to the finer tube.
Further refinement can be achieved through recursion.

Finally, the new layer is evaluated.
First, flux tubes that remain completely outside the first wall are tagged as unnecessary (and are later removed).
For the quality check of the necessary flux tubes, cross-sections are evaluated at $\varphi_k$, and the entire layer is discarded if a non-convex cross-section is detected (there is no need to evaluate unnecessary flux tubes - especially since those can be close to coils and are prone to distortions).
A future improvement could be to repair a (weakly damaged) flux tube by adjusting the base mesh similar to what is proposed in \cite{Takada2021}.
Furthermore, the magnetic flux $\Phi_k$ through each segment $[\varphi_{k-1}, \varphi_k]$ is computed, and the violation of flux conservation

\begin{equation}
\Delta_\Phi \, = \, \frac{\max\Phi_k - \min\Phi_k}{\sum_{k = l+1}^{u} \, \Phi_k} \label{eq:local_flux_violation}
\end{equation}

\noindent is compared to the acceptable error \Dmax provided as input (initially, $l = 0$ and $u = n$).
If the quality check fails, the current toroidal domain is divided in two, and a new base mesh is initialized at the center of each sub-domain as indicated in figure \ref{fig:uqmesh} (c).
The 3D flux tube mesh is constructed recursively from there on radially outward as shown in figure \ref{fig:uqmesh} (d) until the entire volume inside the first wall is filled.

It may be acceptable to allow a few flux tubes with moderate flux conservation error within a layer, in particular if those end up carrying little heat flux.
This is implemented by setting a moderate value for \Dmax and a lower value for \Davg which is provided as additional input parameter.
The latter is used as a constrained for the averaged violation of flux conservation

\begin{equation}
\overline{\Delta}_\Phi \, = \, \frac{1}{\mpnecessary} \, \sum_{i = 1}^{\mpadapt} \, \delta_i \, \Delta_{\Phi \, i}, \qquad
\mpnecessary \, = \, \sum_{i = 1}^{\mpadapt} \, \delta_i \label{eq:average_flux_violation}
\end{equation}

\noindent for the relevant flux tubes within the current layer while $\delta_i = 0,1$ masks contributions from unnecessary ones.
As fewer flux tubes remain towards the first wall, $\overline{\Delta}_\Phi$ from (\ref{eq:average_flux_violation}) is compared to

\begin{equation}
\Davg^\ast \, = \, \Dmax \, + \, \left(\Davg \, - \, \Dmax\right) \, \cdot \, \frac{\mpnecessary - 1}{\mpadapt - 1}.
\end{equation}

\noindent This is equivalent to \Davg for the initial layer but imposes a weaker constraint near the first wall where only a few flux tubes remain. 


\section{Speedup of strike point analysis} \label{sec:examples}

\begin{figure}
\begin{center}
\includegraphics[width=160mm]{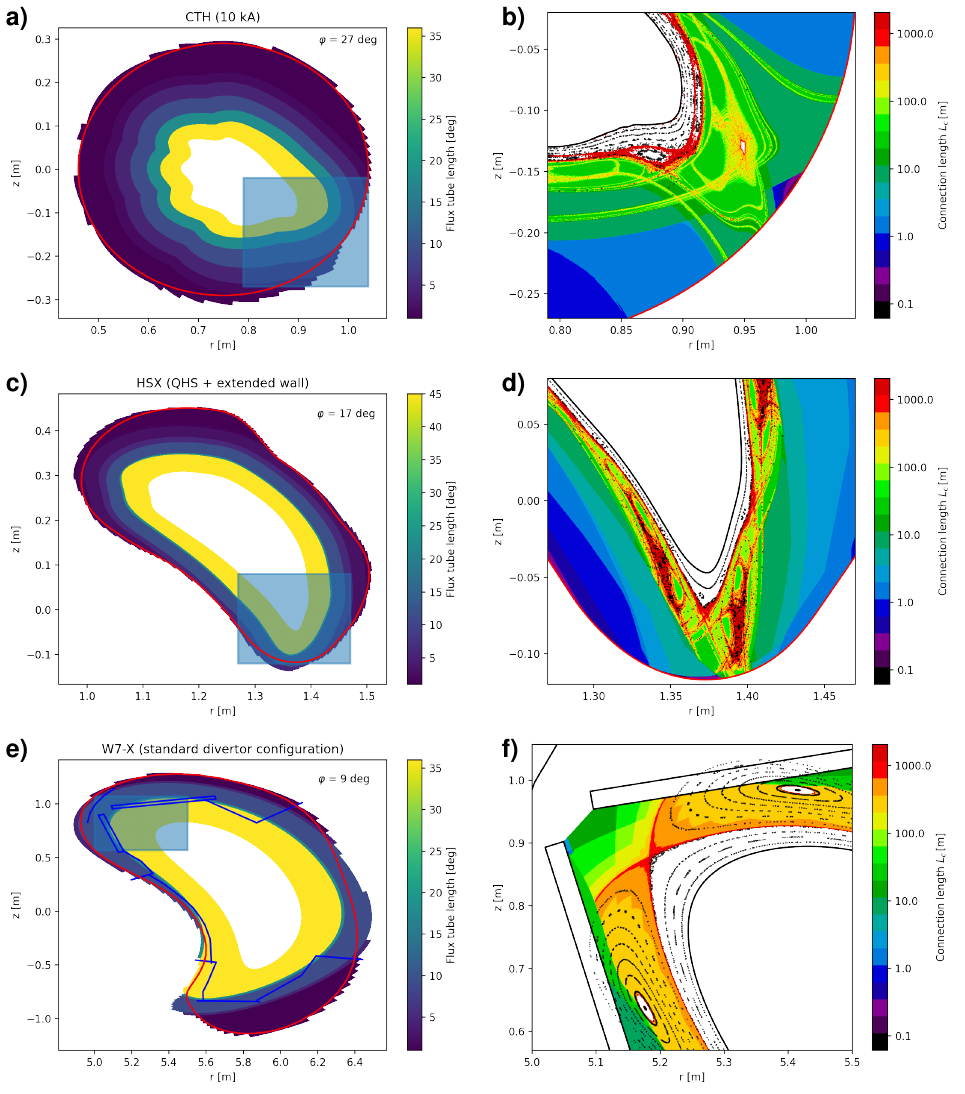}
\caption{Examples for CTH (upper row), HSX (middle row) and W7-X (lower row): (left) flux tube length within mesh, (right) connection length $L_c$ and \Poincare plots from field line reconstruction (divertor targets were deactivated for the \Poincare plot for W7-X in order to highlight the island structure).}
\label{fig:examples}
\end{center}
\end{figure}

The new mesh generator has been applied to configurations at CTH, HSX and W7-X.
The first example shown in the upper row of figure \ref{fig:examples} is the $10 \, \kilo\ampere$ configuration in CTH with a wide chaotic edge \cite{Garcia2023}.
The second example shown in the middle row of figure \ref{fig:examples} is the standard quasi-helically symmetric (QHS) configuration in HSX - but with an extended wall that has been recently proposed \cite{Schmitt2025}.
The third example shown in the lower row of figure \ref{fig:examples} is the standard divertor configuration in W7-X for which only the first wall (red) is used during mesh generation.
Mesh parameters are summarized in table \ref{tab:parameters}.

\begin{table}
\begin{center}\begin{tabular}{l|p{18mm}p{18mm}p{18mm}p{10mm}lll}
   & toroidal symmetry   & major rad. $[\meter]$   & minor rad. $[\meter]$   & $m_p$   & $\Delta_r \, [\milli\meter]$   & \Dmax   & \Davg \\
\hline
CTH    & 5   & 0.75   & 0.29 & 360   & 2    & 0.3    & 0.3 \\
HSX    & 4   & 1.2    & 0.12 & 360   & 3    & 0.3    & 0.3 \\
W7-X   & 5   & 5.5    & 0.53 & 360   & 8    & 0.45   & 0.1
\end{tabular}\end{center}
\caption{Geometry and mesh parameters for the three example configurations in figure \ref{fig:examples}. A toroidal resolution of $0.5 \, \deg$ is used in all cases.}
\label{tab:parameters}
\end{table}

It can be seen in the left column of figure \ref{fig:examples} that flux tubes can be constructed across the entire simulation domain for several layers (yellow) in all configurations before splitting is required.
For W7-X, only far-SOL flux tubes need to be split while almost the entire region of interest with strike points on the divertor targets can be covered with full-length flux tubes.
This is consistent with the traditional EMC3 mesh for this low shear configuration.
By including the far-SOL flux tubes here, the advantage of the new mesh is that it can account for plasma surface interaction along baffles and the entire first wall in W7-X beyond what is typically included in EMC3.
This may be of interest (if integrated into EMC3) in support of material erosion, migration and re-deposition studies for component lifetime analysis for the entire device.

For HSX and CTH, it can be seen that two steps of splitting (cobalt blue) are required for the primary plasma surface interaction area, and that further splitting (dark purple) is required in the far SOL.
Nevertheless, the entire volume within the first wall can be covered with flux tubes.
This is an advantage over the traditional setup of the EMC3-EIRENE mesh for HSX where careful construction of several sub-domains is required already for the much tighter present first wall.
In particular, there is no need to find a suitable outer boundary for the base mesh (at several toroidal locations) which is required in the traditional setup in order to cover relevant areas for plasma surface interaction without any gaps.
Thus, the new mesh is particularly useful for divertor target and baffle design optimization where the primary plasma surface interaction area is subject to change.

One application of the magnetic flux tube mesh is the fast computation of high resolution connection length plots from field line reconstruction.
Examples are shown on the right side of figure \ref{fig:examples}.
For CTH, a fine structured web of field lines in the $100 \, \meter$ range (light green - yellow) is embedded in a layer of field lines in the $10 \, \meter$ range (dark green).
These field lines are expected to carry particles and energy from the last closed flux surface towards the first wall, although cross-field diffusion will likely smooth out the pattern.
Similarly, a fine structured web of long field lines (red) is found in figure \ref{fig:examples} (d) for HSX.
For W7-X, divertor targets and baffles are included now for the connection length computation.
The island separatrix (red) and confined volumes around the O-points (white) are clearly visible in figure \ref{fig:examples} (e).
These plots can be computed 100 - 1000 times faster than with numerical integration of (\ref{eq:ODE}) once a suitable mesh has been constructed.
Nevertheless, due to the overhead from mesh generation, this application is not intended as replacement for the corresponding procedure in FLARE but rather as a supplement for the footprint analysis discussed below.

\begin{figure}
\begin{center}
\includegraphics[width=160mm]{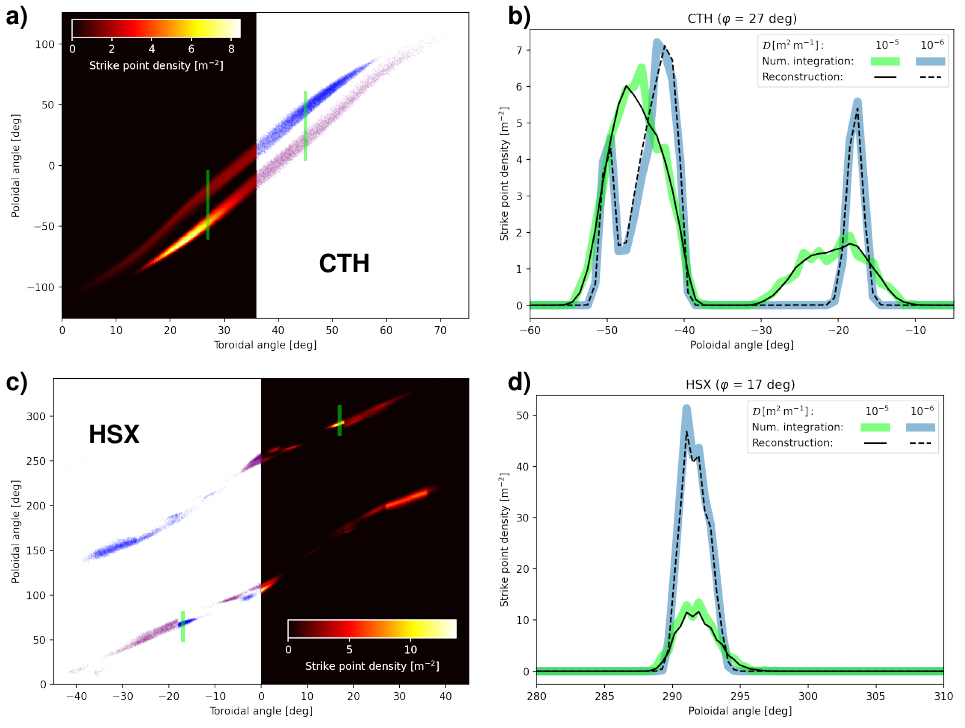}
\caption{Footprint analysis on the first wall for CTH (upper row) and HSX (lower row): (left) strike points in forward (purple) and backward direction (blue) from field line tracing with FLARE (numerical integration) are shown on white background in one half of the field period, and the total (forward + backward) strike point densities computed by field line reconstruction from a magnetic flux tube mesh are shown on black background in the other half of the field period. 
Profiles along the green lines (and their stellarator symmetric counterparts) are shown on the right side and correspond to the cross-sections shown on the right side of figure \ref{fig:examples}.
Additional profiles for smaller field line diffusion ($\mathcal{D} \, = \, 10^{-6} \, \meter^2 \, \meter^{-1}$) are shown as blue and black dashed lines.
}
\label{fig:CTH_and_HSX_footprints}
\end{center}
\end{figure}

\begin{figure}
\begin{center}
\includegraphics[width=160mm]{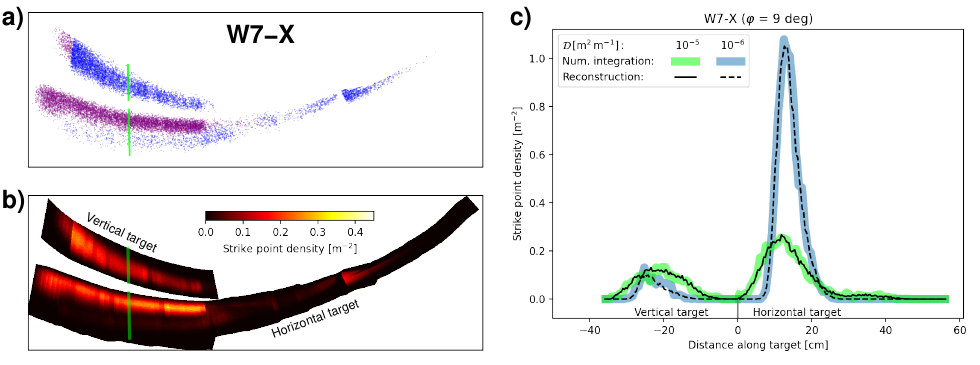}
\caption{Footprint analysis on the divertor targets in W7-X: (a) strike points in forward (purple) and backward direction (blue) from field line tracing with FLARE (numerical integration), (b) the total (forward + backward) strike point density computed by field line reconstruction from a magnetic flux tube mesh, (c) profiles along the green lines from (a) and (b) correspond
to the cross-section shown in figure \ref{fig:examples} (f).
Additional profiles for smaller field line diffusion ($\mathcal{D} \, = \, 10^{-6} \, \meter^2 \, \meter^{-1}$) are shown as blue and black dashed lines.}
\label{fig:W7X_footprint}
\end{center}
\end{figure}

The primary application of the magnetic flux tube mesh is the fast estimation of particle and heat loads on divertor targets for scenario development and optimization.
Artificial cross-field diffusion $\mathcal{D}$ during field line tracing (numerical integration) or reconstruction can mimic particle and energy transport with diffusivity $\chi_\perp$.
The diffusivity $\chi_\perp \, = \, \mathcal{D} \, u$ is related to field line diffusion through a characteristic velocity $u$.
This is implemented as random step

\begin{equation}
\langle\Delta_\perp^2\rangle \, = \, 2 \, \mathcal{D} \, \Delta_\parallel \label{eq:diffusion}
\end{equation}

\noindent for every step $\Delta_\parallel$ along a field line.
For simplicity, (\ref{eq:diffusion}) is taken within the local r-z plane, but a more accurate cross-field step (as in EMC3 \cite{Feng2004}) can be implemented later.
Strike points from numerical integration of field lines with $\mathcal{D} \, = \, 10^{-5} \, \meter^2 \, \meter^{-1}$ are shown in figure \ref{fig:CTH_and_HSX_footprints} (a) and (c) and figure \ref{fig:W7X_footprint} (a) as purple and blue dots.
This value of $\mathcal{D}$ corresponds to $\chi_\perp \, = \, 1 \, \meter^2 \, \second^{-1}$ for $100 \, \electronvolt$ H ions at thermal velocity.
This is more likely to capture the transport upstream near the last closed flux surface than downstream where lower temperatures would correspond to larger $\mathcal{D}$ under the assumption of fixed $\chi_\perp$.
Nevertheless, a small value of $\mathcal{D}$ is of interest for verification of sufficient accuracy of the results from field line reconstruction.
Ultimately, one would want to apply some suitable average value in order to best approximate particle and heat loads.

In order to evaluate the magnetic flux tube mesh, results from field line reconstruction with the same $\mathcal{D}$ are shown as color plots in figure \ref{fig:CTH_and_HSX_footprints} (a) and (c) and figure \ref{fig:W7X_footprint} (b).
Here, instead of the individual strike points, the strike point density $w$ on the surface is shown by counting the number of strike points in each surface cell relative to its area, and which is then normalized to the total count such that $\int\!dA \, w \, = \, 1$.
This can be used as a heat load proxy $q_{t \, (w)} \, = \, w \, \PSOL$ which is proportional to the power \PSOL that needs to be exhausted through the scrape-off layer.
The profiles in figure \ref{fig:CTH_and_HSX_footprints} (b) and (d) and figure \ref{fig:W7X_footprint} (c) show that the strike point densities based on field line reconstruction from the magnetic flux tube mesh (black lines) agree well with those based on numerical integration of field lines (green lines), however, the former computation is 100 - 1000 times faster.
Small differences are expected due to the random nature of the cross-field diffusion step.
The good quality of field line reconstruction is further demonstrated by the same comparison for an even smaller value of $\mathcal{D} \, = \, 10^{-6} \, \meter^2 \, \meter^{-1}$ (blue vs. black dashed lines).

\begin{figure}
\begin{center}
\includegraphics[width=160mm]{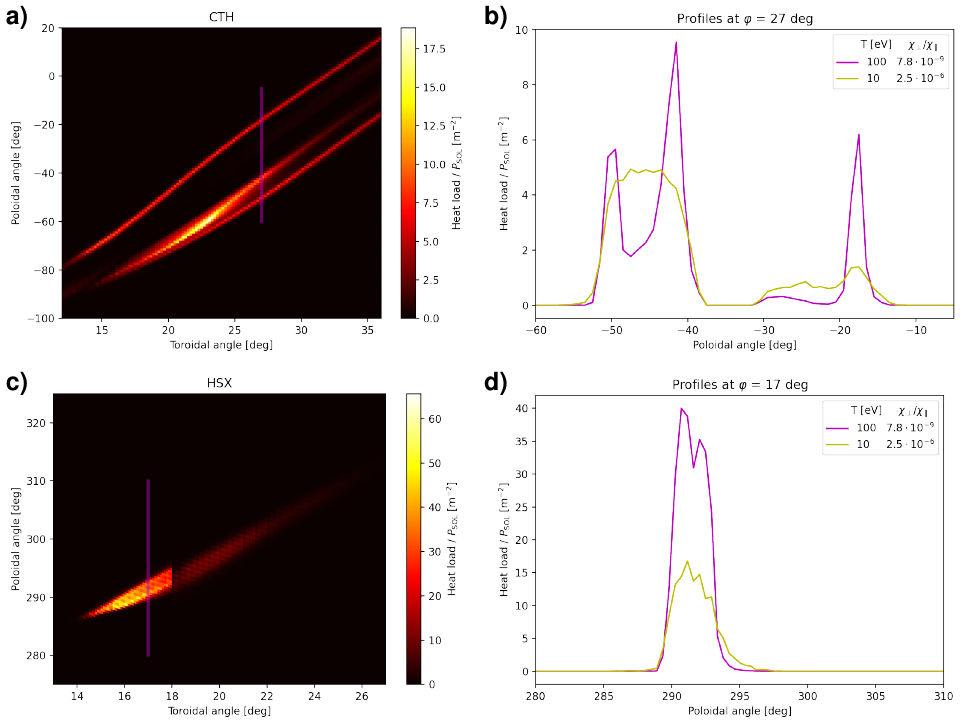}
\caption{Heat load proxy \qproxy / \PSOL for CTH (upper row) and HSX (lower row) based on linearized heat conduction with $\chi_\perp = 1 \, \meter^2 \, \second^{-1}$, $T = 100 \, \electronvolt$ and $n = 10^{19} \, \meter^{-3}$:
profiles along the magenta lines in (a) and (c) are shown in (b) and (d), respectively.
Profiles for $T = 10 \, \electronvolt$ (yellow) with larger $\chi_\perp / \chi_\parallel$ ratio are also shown in (b) and (d) for comparison.
}
\label{fig:CTH_and_HSX_hloads}
\end{center}
\end{figure}

\begin{figure}
\begin{center}
\includegraphics[width=160mm]{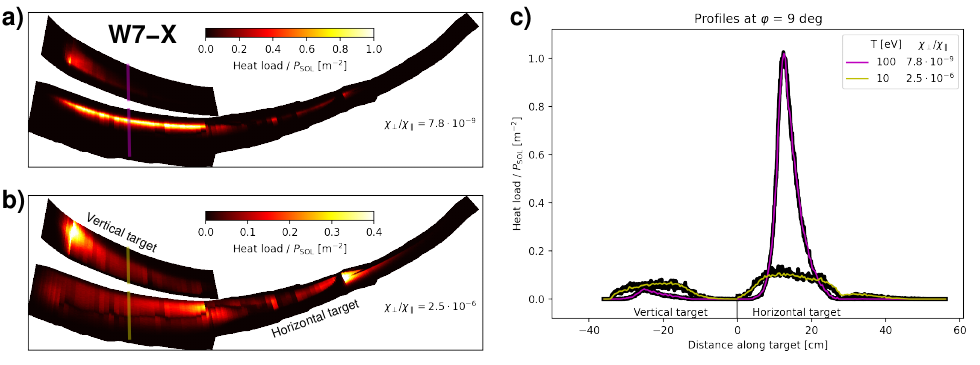}
\caption{Heat load proxy \qproxy / \PSOL for W7-X based on linearized heat conduction with $\chi_\perp = 1 \, \meter^2 \, \second^{-1}$ and $n = 10^{19} \, \meter^{-3}$:
(a) $T = 100 \, \electronvolt$ and (b) $T = 10 \, \electronvolt$.
Profiles at $\varphi = 9 \, \deg$ are shown in (c) in comparison to results from EMC3-Lite (black) based on a structured mesh.}
\label{fig:W7X_hload}
\end{center}
\end{figure}

The heat load proxy $q_{t \, (w)}$ is very similar to the one in EMC3-Lite \cite{Feng2022} in that both are based on alternating steps along field lines and across.
The essential difference, however, is that the step along field lines in EMC3-Lite is based on a linearized version of the heat conduction equation with fixed $\chi_\parallel \, = \, \kappa_\parallel(T) \, / \, n$ based on the heat conductivity $\kappa_\parallel$ for given background density $n$ and temperature $T$.
This implies taking a random step

\begin{equation}
\langle\Delta_\parallel^2\rangle \, = \, 2 \, \chi_\parallel \, \tau \label{eq:linear_heat_conduction}
\end{equation}

\noindent in either direction along a field line rather than taking a fixed step $\Delta_\parallel \, = \, u \, \tau$ while continuing in the same direction.
Even though the latter may result in missing some shadow regions in the far-SOL \cite{Feng2022}, it is still suitable to verify the quality of field line reconstruction against numerical integration.
Based on the verification presented here, a linearized heat conduction step (\ref{eq:linear_heat_conduction}) is also implemented.
The corresponding cross-field step $\langle\Delta_\perp^2\rangle \, = \, 2 \, \chi_\perp \, \tau$ is equivalent to (\ref{eq:diffusion}) for field line diffusion when the time step $\tau$ is used instead of the space step $\Delta_\parallel$.
The resulting heat load proxy \qproxy is shown in figures \ref{fig:CTH_and_HSX_hloads} and \ref{fig:W7X_hload} for the same three example configurations as before.
Cross field transport is set to $\chi \, = \, 1 \, \meter^2 \, \second^{-1}$, and parallel transport is computed from $n = 10^{19} \, \meter^{-3}$ and $T = 100 \, \electronvolt$ which results in the small ratio of $\chi_\perp / \chi_\parallel \, = \, 7.8 \, \cdot \, 10^{-9}$ due to the strong temperature dependence of $\kappa_\parallel \, \sim \, T^{5/2}$.
This results in very peaked heat loads.
In particular, the thin bands of long connection length field lines shown in figure \ref{fig:examples} (b) for CTH are recovered.
The profiles (magenta) in figures \ref{fig:CTH_and_HSX_hloads} (b) and (d) and \ref{fig:W7X_hload} (c) resemble the ones for the strike point density at $\mathcal{D} \, = \, 10^{-6} \, \meter^2 \, \meter^{-1}$ in figures \ref{fig:CTH_and_HSX_footprints} (b) and (d) and figure \ref{fig:W7X_footprint} (c), respectively.
In particular, figure \ref{fig:W7X_hload} (c) shows that results are in very good agreement with EMC3-Lite (black profiles below colored ones).
Computationally, the structured mesh layout of EMC3-Lite may allow for a somewhat higher efficiency than the unstructured mesh layout of the new approach.
The new approach, however, allows for greater configurational flexibility - particularly for high shear configurations - with an automated procedure where meticulous manual fine tuning may be required otherwise.

As pointed out earlier, the high temperature assumption likely underestimates the impact of cross-field diffusion downstream - even more so for the heat conduction step (\ref{eq:linear_heat_conduction}).
Results for $T = 10 \, \electronvolt$, on the other hand, are shown as yellow profiles in figures \ref{fig:CTH_and_HSX_hloads} (b) and (d) and \ref{fig:W7X_hload} (c), and those are in line with the strike point density for $\mathcal{D} \, = \, 10^{-5} \, \meter^2 \, \meter^{-1}$.
This underlines the importance of choosing a reasonable $T$ (as well as $\chi_\perp$ and $n$) for a good heat load proxy here as well as in EMC3-Lite.
It may be prudent to explore a range of different $\chi_\perp / \chi_\parallel$ values when applied for optimization of baffles and divertor targets.


\section{Conclusions}

A new automated mesh generator for unstructured quadrilateral flux tubes with adaptive refinement has been presented as an extension of the FLARE code.
It has been found that several layers of flux tubes can span the entire half field period before splitting of flux tubes is required.
This is an advantage over the traditional setup of the EMC3-EIRENE mesh where careful construction of several sub-domains is required for configurations like HSX.
In particular, there is no longer the need to find a suitable outer boundary for the base mesh (at several toroidal locations).
The strike point density pattern from field line reconstruction agrees well with the corresponding one from numerical integration, but computations are 100 - 1000 times faster.
Thus, the new adaptive flux tube mesh generator combined with fast head load approximation offers a powerful tool for integration into optimization workflows for divertor target and baffle design.
An extension of EMC3-EIRENE for more flexible mesh layouts may be of interest as a future project.


\appendix
\section*{Acknowledgments}
This work was supported by the U.S. Department of Energy under awards No. P-240001537, DE-SC0014210 and DE-FG02-93ER54222.
One of the authors was also supported by the Advanced Opportunity Fellowship (AOF) through the Graduate Engineering Research Scholars (GERS) at University of Wisconsin-Madison.

\section*{References}


\end{document}